\documentclass[conference]{IEEEtran}
\IEEEoverridecommandlockouts

\usepackage{cite}
\usepackage{amsmath,amssymb,amsfonts}
\usepackage{bm}
\usepackage{algorithmic}
\usepackage{graphicx}
\usepackage{textcomp}
\usepackage{xcolor}
\usepackage{subcaption}
\usepackage{algorithm}
\usepackage{algorithmic}
\usepackage{comment}
\usepackage{caption}
\usepackage[
    bookmarks=false,
    draft
]{hyperref}
\def\BibTeX{{\rm B\kern-.05em{\sc i\kern-.025em b}\kern-.08em
    T\kern-.1667em\lower.7ex\hbox{E}\kern-.125emX}}
\begin{document}

\title{Movable Antenna Optimization for Multi-User MIMO Systems in Realistic Ray-Traced Propagation Environments}

\author{Xiaoyi~Zhang,~Amna~Ishad, and Emil~Bj{\"o}rnson\\
\IEEEauthorblockA{\textit{KTH Royal Institute of Technology, Stockholm, Sweden}\\  Email: \{xiaoyiz,amnai,emilbjo\}@kth.se}

}

\maketitle

\begin{abstract}
To meet the growing data traffic demand in future wireless systems, novel transmission architectures capable of adapting to complex propagation environments are required. Movable antenna (MA) systems have recently emerged as a promising approach, enabling the physical repositioning of antenna elements to exploit spatial degrees of freedom. However, existing studies largely rely on idealized or simplistic channel models, leaving open the question of whether the performance gains of MA systems persist under realistic propagation conditions.
This paper investigates the performance of downlink multi-user MIMO systems with movable antennas using deterministic ray-traced channel models. A simulation framework combining three-dimensional ray tracing and field-response channel modeling is developed, and antenna positions are optimized using particle swarm optimization and genetic algorithms. Simulation results reveal that while simplified distance-based channel models predict large performance disparities between competing array configurations, realistic ray-traced channels significantly compress these differences, indicating that propagation effects dominate over pure array geometry optimization. Nevertheless, movable antenna systems retain strong effectiveness over conventional fixed arrays across different user distributions, array sizes, and multipath conditions, even in geometry-constrained propagation environments.
\end{abstract}


\section{Introduction}
The ever-increasing demand for higher data rates, lower latency, and more flexible spatial coverage has been a primary driver behind the evolution of wireless communication systems, from 4G to 5G and now to the forthcoming 6G networks. Meeting these requirements challenges conventional transmission architectures and has motivated the investigation of new designs that extend beyond traditional fixed-antenna configurations at base stations (BSs). While BS locations remain static in most deployments, recent advances enable antenna elements at the BS to be physically reconfigured, opening new possibilities for spatial adaptation.

Among emerging paradigms, movable (via mechanical devices like stepper motors) or fluid (port selection ~\cite{Kit21fluid,WEEfluid24}) antenna (MA/FAS) systems have attracted growing attention due to their ability to dynamically adjust antenna positions in real time 
~\cite{ZhuLipeng2023MAfW}. Related implementations, such as flexible intelligent meta-surfaces \cite{FIM25}  or reconfigurable pixel antennas \cite{REMAA25}, rely on different physical mechanisms but share the same principle of spatial antenna adaptation considered in this work. By adapting antenna positions to user locations and propagation conditions, MA/FAS systems can enhance the channel matrix structure, for example, by strengthening dominant singular modes in single-user scenarios or by improving channel orthogonality across users in multi-user settings. These effects can translate into higher spectral efficiency and more robust link performance, especially in propagation environments where conventional fixed arrays face challenging interference and fading issues. In practice, mechanical antenna repositioning occurs on a slower time scale than small-scale fading, making MA systems particularly suitable for quasi-static scenarios where antenna positions can be periodically optimized based on slowly varying channel conditions.

Extensive research has been conducted on MA/FAS optimization. Antenna positions, receive filters, and power allocation were jointly optimized in~\cite{xiao2024multiuser} to maximize the minimum user rate, while this framework was extended in~\cite{gao2023mimo} to MIMO systems to enhance channel capacity. Power-efficient and trajectory-based approaches were explored in~\cite{wu2023movable,zhang2022graph}, and alternating optimization and weighted minimum mean square error methods were applied in~\cite{liu2023joint,wang2023wmmse} for signal-to-interference-plus-noise ratio and weighted rate maximization, respectively. The authors in \cite{yan2025new} also tested their MA optimization on statistical channel state information using ray tracing in Singapore. Collectively, these works demonstrate the potential of antenna mobility to improve system throughput by leveraging favorable channel realizations.

Accurate channel modeling is essential for reliably assessing the performance of wireless systems. Ray tracing has been established as a powerful deterministic approach for modeling site-specific propagation effects, including reflection, diffraction, and scattering. Early geometry-based models~\cite{saleh1987statistical,seidel1994site} demonstrated high accuracy in localized environments, while more recent developments enable efficient three-dimensional ray-tracing simulations for large-scale MIMO systems~\cite{liu2020raytracing3d} and hybrid deterministic data-driven models~\cite{chen2021raytracing}. Ray tracing has also been successfully applied to emerging scenarios such as UAV-assisted communications and sub-terahertz systems~\cite{li2021raytracinguav,yun2015raytracing}, underscoring its relevance for guiding next-generation wireless design before large-scale testbeds are built.

Based on these foundations, this work investigates whether the performance gains of movable antenna systems persist when using deterministic, ray-traced channel models in complex three-dimensional propagation environments. To this end, we provide an empirical and insight-driven study of MA systems using realistic urban ray-tracing simulations, rather than the simplified distance-based channel models commonly adopted in the literature. Specifically, our contributions are: 
\begin{itemize}
    \item We show through extensive ray-tracing simulations that the performance gains of MA systems persist under deterministic ray-tracing channels, despite the reduced channel richness and strong geometric constraints.
    \item We identify a fundamental trade-off between antenna mobility and array aperture, showing that the benefits of MA diminish as fixed-array size increases, providing practical guidance on when MA architectures are most beneficial.
    \item We show that MA systems can effectively exploit dominant specular propagation paths in ray-traced environments, yielding robust performance gains across varying levels of multipath richness. 
\end{itemize}

\section{System Model}

\begin{figure*}[t!]
\centering
\subfloat[BS with movable antenna elements serving multiple users.] 
{\includegraphics[width=0.4\textwidth]{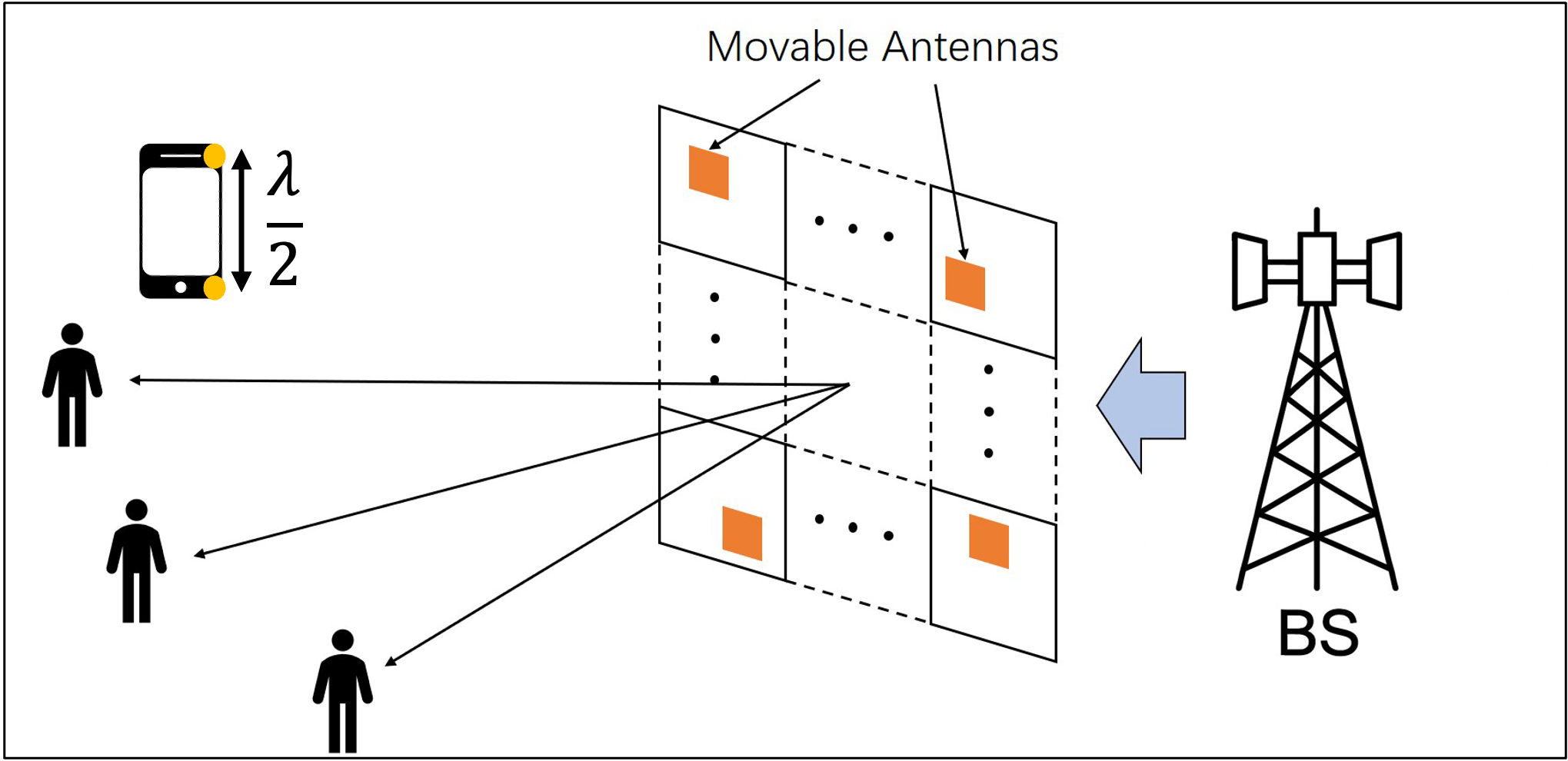}}\hspace{1cm}
\subfloat[Antenna elements in a 3D coordinate system.]
{\includegraphics[width=0.4\textwidth]{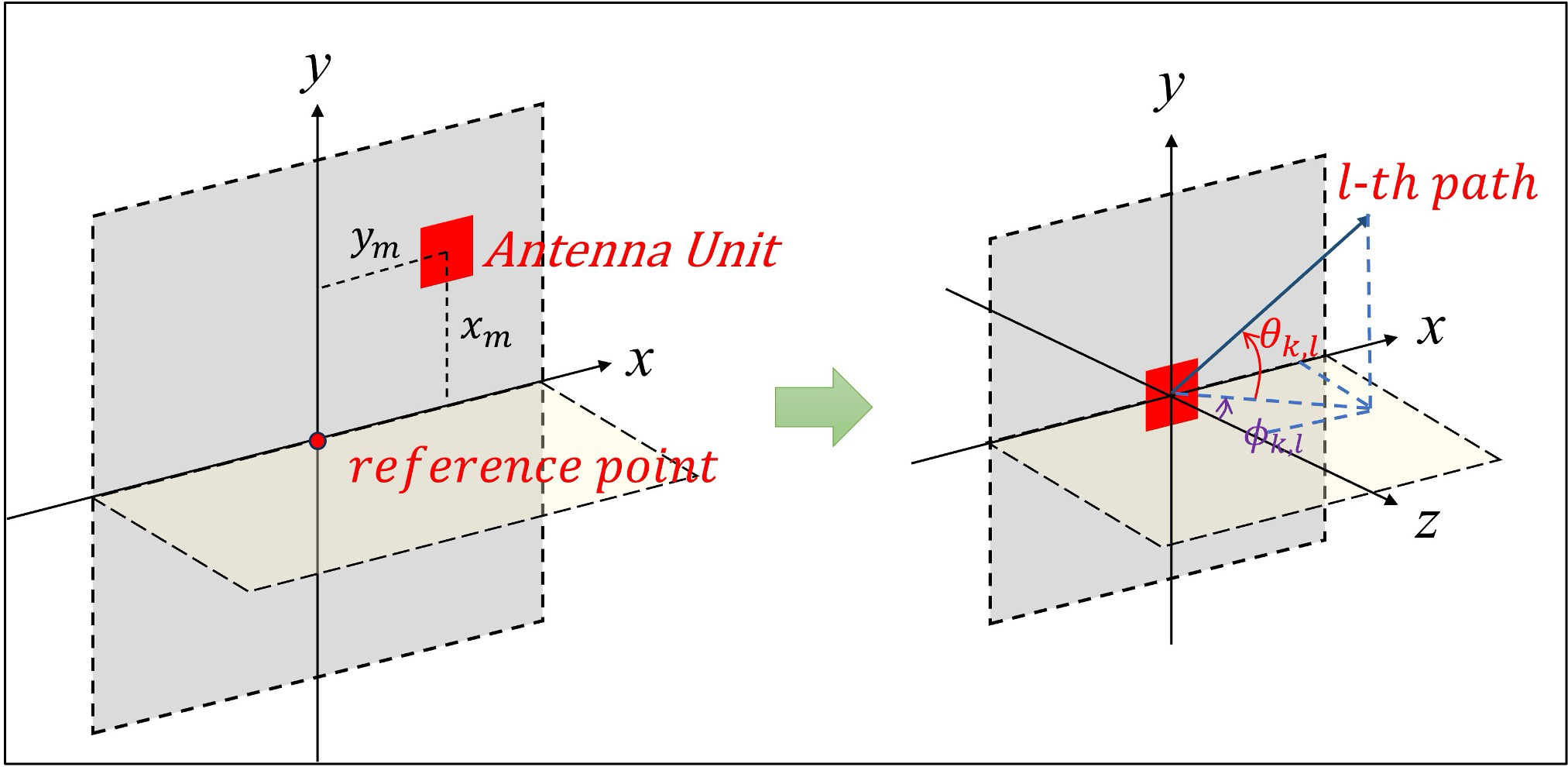}}
\centering
\caption{Depiction of the system setup.}
\label{fig:system}
\end{figure*}

We consider a downlink multi-user MIMO system, where a base station (BS) with $M$ MA elements is communicating with $N$ users. The $m$-th MA can move within a region $\mathcal{C}_m$ in the x-y plane.
Each user is equipped with $N_r$ receive antennas, forming a uniform linear array (ULA), as depicted in Fig.~\ref{fig:system}(a). 
 To characterize the propagation between the BS and the users, we adopt the general field response channel model from~\cite{xiao2024multiuser}, which expresses the channel coefficients as an explicit function of antenna positions and multipath parameters. This formulation enables fine-grained spatial control of the array response by adjusting antenna locations.
In this work, the multipath parameters required by this model, including path loss, angles of departure (azimuth and elevation), and the number of propagation paths, are obtained using a geometry-based ray tracing framework. Specifically, line-of-sight (LoS) components are determined via direct ray tracing, while non-line-of-sight (NLoS) paths are generated using the Shooting and Bouncing Rays (SBR) method~\cite{schaubach1992ray}.
The SBR method simulates signal propagation by launching a large number of rays from the transmitter across a geodesic sphere in nearly uniform directions, capturing diverse interactions with the surrounding environment. As rays propagate, they may encounter surfaces and edges, leading to reflections and diffraction that collectively characterize the multipath channel.

As shown in Fig.~\ref{fig:system}(b), for the $n$-th antenna of $i$-th user and its $l$-th propagation path, the phase shift at a movable antenna located at position $\mathbf{r}_m = [x_m, y_m]^T$ is
\begin{equation}
\rho_{i,n,l}(\mathbf{r}_m) = x_m \sin(\theta_{i,l}) \cos(\phi_{i,l}) + y_m \cos(\theta_{i,l}),
\end{equation}
where $\theta_{i,l}$ and $\phi_{i,l}$ denote the elevation angle measured from the horizontal plane and azimuth angle measured in the x-y plane, respectively. Based on this, the field response vector at the $m$-th BS antenna for the $i$-th user is defined as
\begin{equation}
\mathbf{f}_{i,n}(\mathbf{r}_m) = 
\begin{bmatrix}
e^{j\frac{2\pi}{\lambda} \rho_{i,n,1}(\mathbf{r}_m)} \\
\vdots \\
e^{j\frac{2\pi}{\lambda} \rho_{i,n,L_i}(\mathbf{r}_m)}
\end{bmatrix},
\end{equation}
where $\lambda$ is the carrier wavelength and $L_i$ is the number of propagation paths. The overall channel vector between the MA array and the $i$-th user is then computed as
\begin{equation}
\mathbf{h}_{i,n}(\tilde{\mathbf{r}}) = \mathbf{F}_i^H(\tilde{\mathbf{r}})\mathbf{g}_i,
\end{equation}
where $\mathbf{F}_i(\tilde{\mathbf{r}}) = [\mathbf{f}_i(\mathbf{r}_1), \dots, \mathbf{f}_i(\mathbf{r}_M)] $ is the \textit{field response matrix} for the $i$-th user, constructed by stacking the responses across all $M$ BS antennas, and $\mathbf{g}_i $ contains the \textit{path loss} obtained from the ray tracing. The overall channel matrix from the BS to $i$-th user can be expressed as $\mathbf{H}_i = [\mathbf{h}_{i,1}^T(\tilde{\mathbf{r}}), \dots,\mathbf{h}_{i,N_r}^T(\tilde{\mathbf{r}})]\in \mathbb{C}^{N_r \times M}$ and the total channel matrix becomes $\mathbf{H} = [\mathbf{H}_1^T, \dots, \mathbf{H}_N^T]^T$. Based on the above channel model, we now formulate the downlink multi-user precoding problem.

\section{Problem Formulation}

Let $\mathbf{d}_i \in \mathbb{C}^{d_i}$ denote the data vector intended for the $i$-th user and $\mathbf{W}_i\in \mathbb{C}^{M \times d_i}$ be the corresponding precoding matrix, then the received signal at the $i$-th user is
\begin{equation}
    \mathbf{y}_i = \mathbf{H}_i \mathbf{W}_i \mathbf{d}_i + \sum_{j=1, j\neq i}^{N} \mathbf{H}_i \mathbf{W}_j \mathbf{d}_j + \mathbf{n},
\end{equation}
where $\mathbf{n} \sim \mathcal{CN}(\mathbf{0}, \sigma^2 \mathbf{I})$ is additive white Gaussian noise and $\sigma^2$ is the noise power. The precoding matrix $\mathbf{W}_i$ is calculated via Block Diagonalization (BD), such that the signal intended for the $i$-th user lies in the null space of the channels of all other users, thereby achieving zero inter-user interference under ideal conditions \cite{spencer}. BD is known to be most effective in moderate-to-high SNR regimes. While suboptimal in general settings, BD is adopted here as a baseline linear precoding strategy to isolate the impact of antenna position optimization under realistic channel conditions. Since the same precoding scheme is used consistently across all compared configurations, the observed performance gains can be attributed primarily to the movable antenna deployment rather than the choice of precoder. For single-antenna users, BD reduces to zero-forcing (ZF) precoding. Specifically, BD is implemented by constructing the null space of the aggregated channel matrix of all users except the intended $i$-th user:
\begin{equation}
    \bar{\mathbf{H}}_i = \left[\mathbf{H}_1^T, \dots, \mathbf{H}_{i-1}^T, \mathbf{H}_{i+1}^T, \dots, \mathbf{H}_N^T\right]^T.
\end{equation}
Let $\bar{\mathbf{V}}_i^{(0)}$ be an orthonormal basis for the null space of $\bar{\mathbf{H}}_i$, i.e., $\bar{\mathbf{H}}_ i\bar{\mathbf{V}}_i^{(0)} = \mathbf{0}$. Then, the effective channel becomes 
\begin{equation}
    \mathbf{H}_i \bar{\mathbf{V}}_i^{(0)} = \mathbf{U}_i 
    \begin{bmatrix}
        \bm{\Sigma}_i & \mathbf{0} \\
        \mathbf{0} & \mathbf{0}
    \end{bmatrix}
    \begin{bmatrix}
        \mathbf{V}_i^{(1)} \mathbf{V}_i^{(0)}
    \end{bmatrix}^H.
\end{equation}
The BD precoder is defined as
\begin{equation}
    \mathbf{W}_i = \bar{\mathbf{V}}_i^{(0)} \mathbf{V}_i^{(1)} \mathbf{D}_i^{1/2},
\end{equation}
where $\mathbf{V}_i^{(1)}$ contains the right singular vectors corresponding to non-zero singular values of the effective channel, and $\mathbf{D}_i$ is the diagonal power allocation matrix and is computed using the water-filling method \cite{bjornson2024introduction} to maximize the sum rate under the constraint that the total transmit power of the BS is $P$. Then the sum rate of $i$-th user is computed as
\begin{equation}\label{sumrate}
    R_i = \log_2 \det\left( \mathbf{I} + \frac{1}{\sigma^2} \mathbf{H}_i \mathbf{W}_i \mathbf{W}_i^H \mathbf{H}_i^H \right).
\end{equation}
Our goal is to optimize the sum rate of the system and define the optimization problem as
\begin{align}
\underset{\mathbf{r}_1,\ldots,\mathbf{r}_M}{\mathrm{maximize}} \quad & \sum_{i=1}^{N} R_i,  \\
\text{subject to} \quad & \mathbf{r}_m \in \mathcal{C}_m, \quad 1 \leq m \leq M,  \\
& \| \mathbf{r}_i - \mathbf{r}_j \| \geq \lambda / 2, \quad 1 \leq i \neq j \leq M,
\end{align}
where the first constraint represents that each antenna element can only move within a corresponding confined range and the second constraint defines a minimum spacing between each antenna element to be greater than $\lambda/2$, to have minimal mutual coupling between antennas \cite{Yuan2023a}.

\section{Proposed Solution Approaches}

To optimize the sum rate in \eqref{sumrate}, two population-based optimization algorithms, namely particle swarm optimization (PSO)~\cite{liu2025time} and genetic algorithm (GA) \cite{song14gen}, are employed and compared with benchmark schemes. These algorithms are particularly suitable for the considered problem due to its non-convexity, high dimensionality, and nonlinear geometric constraints. Population-based algorithms, such as PSO and GA, provide an effective trade-off between solution quality and computational complexity for such problems, enabling near-optimal antenna placements without requiring gradient information.

\subsection{Particle Swarm Optimization (PSO)}

In PSO, each particle represents a candidate solution to the antenna placement problem. In our case, a particle is encoded as a vector of $2M$ real-valued variables corresponding to the $(x,y)$ coordinates of the $M$ movable antennas at the BS. The swarm (all particles) iteratively updates particle positions and velocities based on both the particle’s individual best known position and the global best position found by the swarm.

To enforce the minimum inter-element spacing constraint, a penalty-based approach is adopted. Configurations in which the distance between any two antennas is smaller than half the wavelength $\lambda/2$ are penalized during the fitness evaluation. The resulting penalized fitness function is defined as
\begin{equation}
f =  \sum_{i=1}^{N} R_i - \alpha \sum_{i\neq j} \mathbb{I}\left( \| \mathbf{r}_i - \mathbf{r}_j \|  <\frac{\lambda}{2} \right),
\end{equation}
where $\mathbf{r}_i$ and $\mathbf{r}_j$ denote the BS positions of the $i$-th and $j$-th antennas, respectively, $\alpha$ is a penalty coefficient, and $\mathbb{I}(\cdot)$ is an indicator function that equals one when the constraint is violated and zero otherwise. This penalty mechanism discourages infeasible antenna configurations while preserving the flexibility of the search process.
 
\subsection{Genetic Algorithm (GA)} 
GA follows a similar solution encoding strategy, where each individual (chromosome) in the population is represented by a vector of $2M$ real-valued variables corresponding to the antenna coordinates. GA employs tournament selection~\cite{kanade2023genetic} to choose parent chromosomes, whereby a subset of individuals is randomly selected and the fittest among them is retained for reproduction. Crossover and mutation operations are then applied to generate new offspring solutions. This process proceeds for a fixed number of generations or until a convergence criterion is satisfied. The individual with the highest fitness value in the final population is selected as the optimized antenna placement solution. The same penalized fitness function used in PSO is adopted for GA, ensuring a fair and consistent comparison between the two algorithms.

\section{Numerical Results}
\begin{figure*}[t!]
\centering
\subfloat[3D view of BS (red) serving two users (blue).] 
{\includegraphics[width=0.25\textwidth]{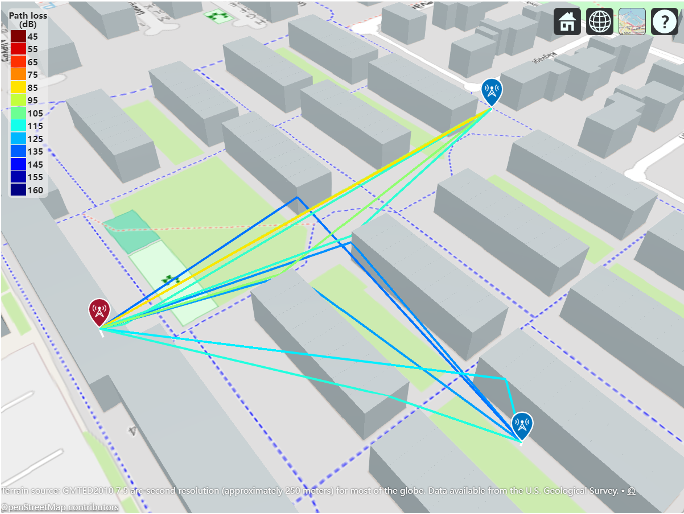}}\hspace{1cm}
\subfloat[The user distribution range.]
{\includegraphics[width=0.3\textwidth]{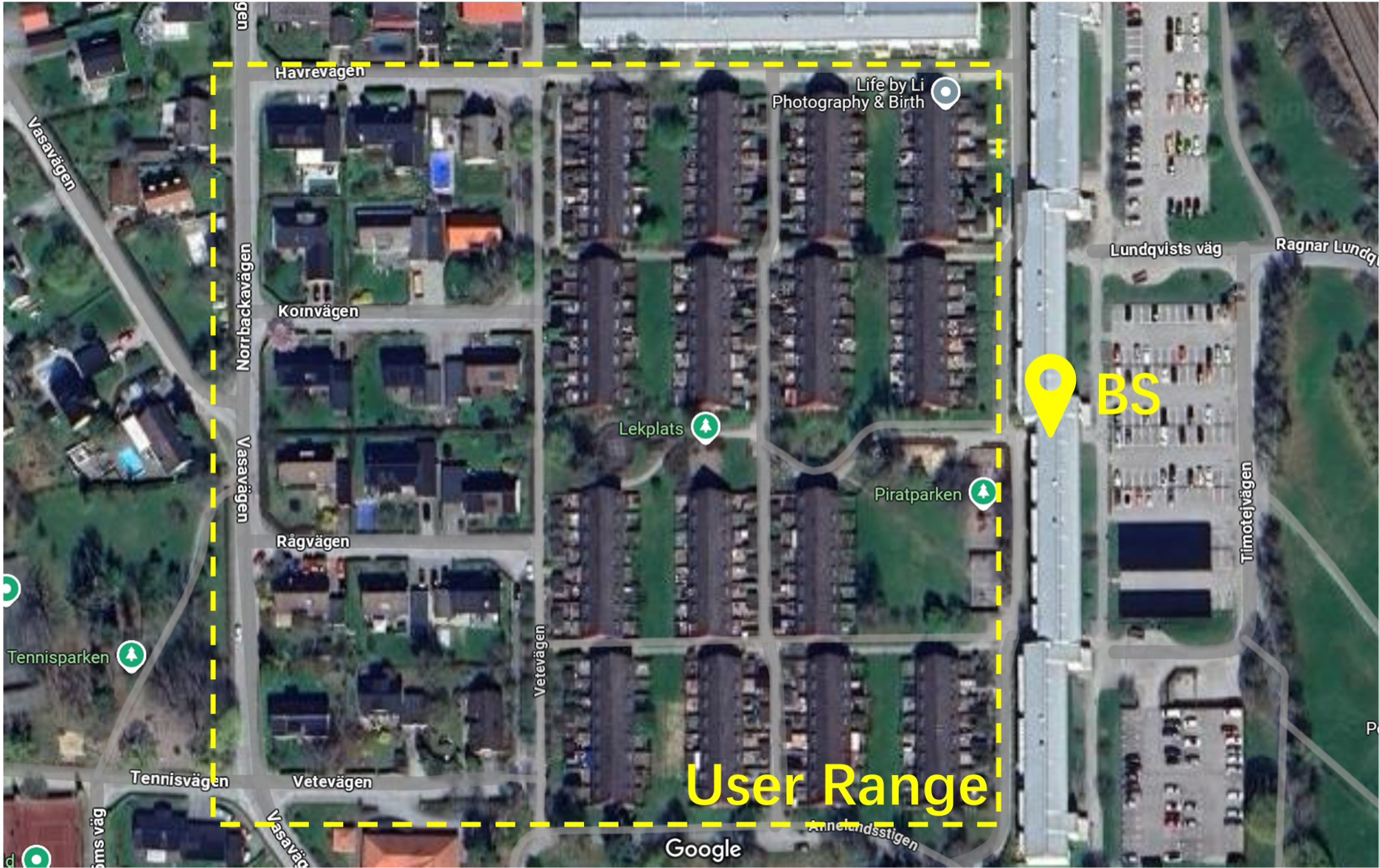}}
\centering
\caption{Ray tracing simulation setup.}
\label{fig:rt_simulation}
\end{figure*}
We evaluate the system performance using the parameter settings summarized in Table~\ref{tab:sim_parameters}, consistent with typical small-cell base station deployments. Unless otherwise stated, the ray-tracing parameters (reflection coefficients, material properties, and path-loss models) follow the default settings of MATLAB’s ray-tracing toolbox, ensuring reproducibility of the results. The considered simulation environment is shown in Fig.~\ref{fig:rt_simulation}, where the red marker denotes the BS and blue markers represent the users. Due to the presence of buildings between the BS and the users, both LoS and NLoS propagation components are present, thereby closely mimicking a realistic urban environment. A total of five users with two antennas each (separated by $\lambda/2$) are considered. 

The number of iterations (200) and population size (100) were selected to ensure convergence of both PSO and GA, as preliminary simulations showed negligible performance improvement beyond 200 iterations. All reported results shown in this section correspond to the converged solutions.

\begin{table}[t!]
\centering
\caption{Simulation Parameters}
\begin{tabular}{l|l|l|l}
\hline
\textbf{Parameter} & \textbf{Value} & \textbf{Parameter} & \textbf{Value}\\
\hline
Base station height & $3$ m & Transmit power of BS & $37$ dBm\\
\hline
Number of users & $5$ & Antennas per user & $2$\\
\hline
User height & $1.5$ m & Carrier frequency & $3$ GHz\\
\hline
Bandwidth & $100$ MHz & Noise power & $-84$ dBm$\!\!\!$\\
\hline
Number of reflection & [1,3,5,10] & Number of diffraction & $1$\\
\hline
\end{tabular}
\label{tab:sim_parameters}
\end{table}

In addition to MA, traditional fixed BS schemes are also considered. The overall benchmark schemes are:

\begin{itemize}

    \item \textbf{Interference-free bound (Inter-free bound/ Upper bound)}: BS equipped with an MA array, where an interference-free upper bound assuming orthogonal user channels.
    \item \textbf{Proposed MA-PSO}: The BS equipped with an MA array, where the antenna positions are optimized using PSO.

    \item \textbf{Proposed MA-GA}: The BS with an MA array where the antenna positions are optimized using GA.

    \item \textbf{Sparse Planar Array (F-SPA)}: A fixed sparse planar array with an inter-element spacing  of $5\lambda$.
        
    \item \textbf{Half-wavelength Planar Array (F-HwPA)}: A conventional fixed planar array with a $\lambda/2$-element spacing.
    
\end{itemize}

 For MA configurations, the antenna elements are allowed to move within a 5$\lambda$×5$\lambda$ square region. For a fair comparison, all considered schemes employ the same total number of BS antennas, i.e., $M=16$. For each configuration, users are randomly placed according to a spatially uniform distribution over a squared region centered at the BS, and performance metrics are evaluated over multiple independent realizations. Both the average sum rate and the empirical CDFs are reported to capture the statistical behavior induced by the randomness of the user location. We first evaluate whether the performance gains of movable antenna systems persist under deterministic ray-tracing (RT) channels, which capture realistic multipath propagation and geometric blockage effects. To highlight the impact of channel modeling assumptions, we additionally consider a simplified distance-based channel model, where each transmit receive antenna pair has a single propagation path with distance-dependent path loss and phase determined by the propagation delay, while neglecting multipath, angular dispersion, and blockage.

\begin{figure}[t!]
    \centering
     \includegraphics[width=0.5\textwidth]{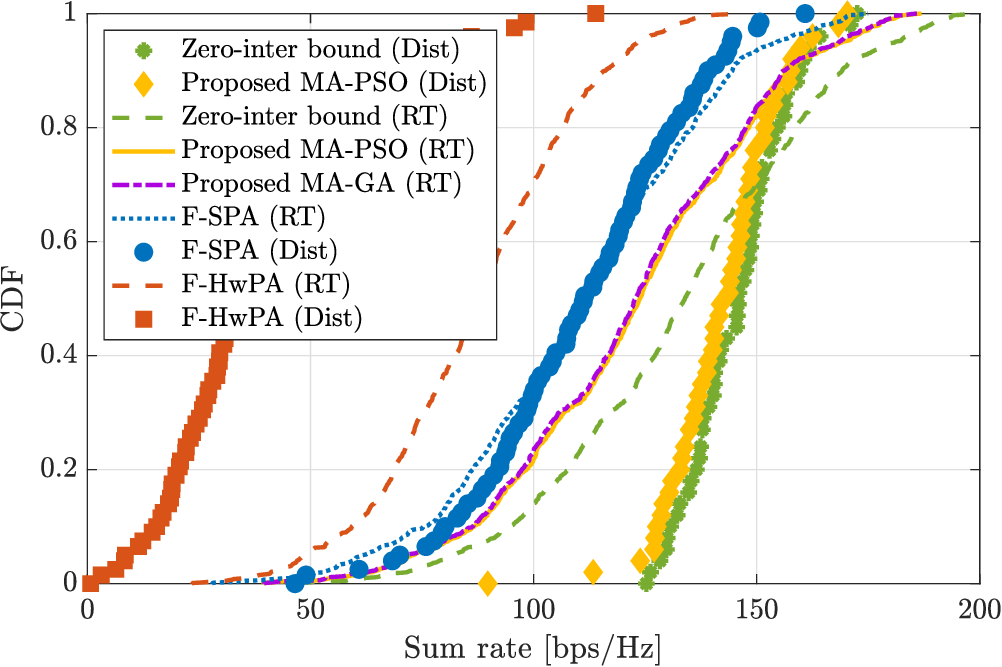}
    
    \caption{CDF of sum rate for distance-based vs. ray-traced channel model.}
    \label{fig:result_cdf}
\end{figure}
Fig.~\ref{fig:result_cdf} shows the empirical CDFs of the downlink sum rate for movable and fixed antenna configurations under both distance-based and ray-traced channel models. Large performance gaps are observed under the distance-based model, whereas ray-traced channels significantly compress these differences, indicating that realistic propagation effects dominate over array geometry optimization. Nevertheless, the proposed MA schemes achieve approximately 92\% of the interference-free benchmark under RT channels, demonstrating that antenna mobility remains effective even in realistic geometry-based environments. The performance difference between PSO and GA is negligible, indicating that both optimization algorithms converge to similar high-quality antenna configurations. Among fixed array benchmarks, F-SPA consistently outperforms F-HwPA. Although the compact $\lambda/2$ spacing of F-HwPA avoids grating lobes, it limits the effective array aperture and achievable array gain, resulting in the lowest performance. In contrast, the larger aperture of F-SPA compensates for the potential grating-lobe effect in the considered environment, leading to improved sum-rate performance.

\begin{figure}[t!]
    \centering
    \begin{subfigure}[b]{0.15\textwidth}
        \centering
        \includegraphics[width=\textwidth]{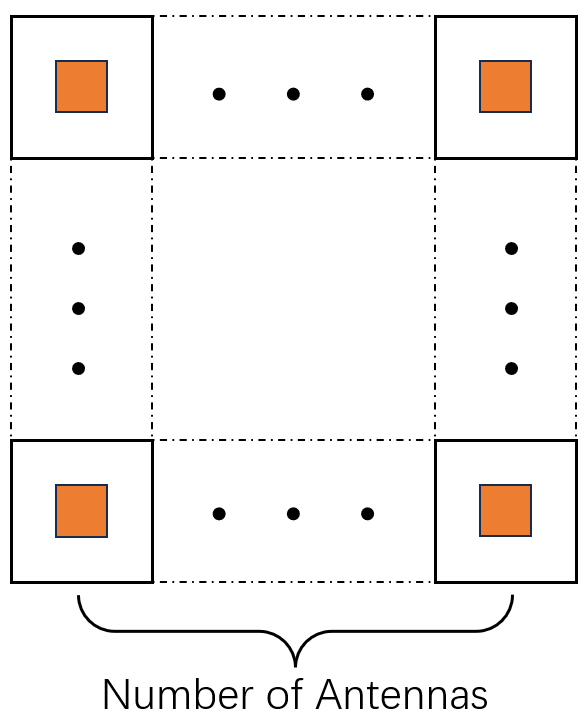}
    \end{subfigure}
    \begin{subfigure}[b]{0.33\textwidth}
        \centering
        \includegraphics[width=\textwidth]{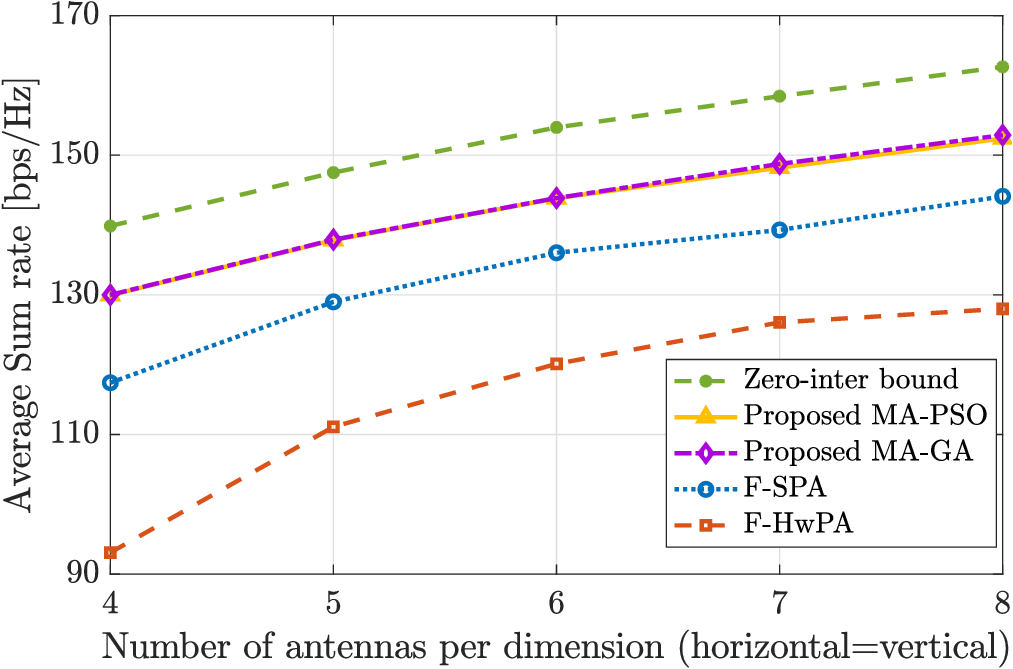}
    \end{subfigure}
    \caption{Sum rate for different number of BS antennas.}
   \label{fig:result_numAntenna}
\end{figure}

Fig.~\ref{fig:result_numAntenna} depicts the sum rate performance as a function of the number of BS antennas. The proposed MA schemes optimized by PSO and GA consistently outperform the fixed SPA and HwPA configurations across all array sizes. This demonstrates that antenna position optimization can significantly enhance system performance, although at the cost of continuous antenna movement. Moreover, as the array size increases, the performance gap between SPA and MA-based schemes diminishes, indicating that large fixed arrays can partially compensate for the lack of antenna mobility. This suggests that the primary benefit of MA lies in scenarios where physical array size is constrained.

\begin{figure}[t!]
    \centering
     \includegraphics[width=0.45\textwidth]{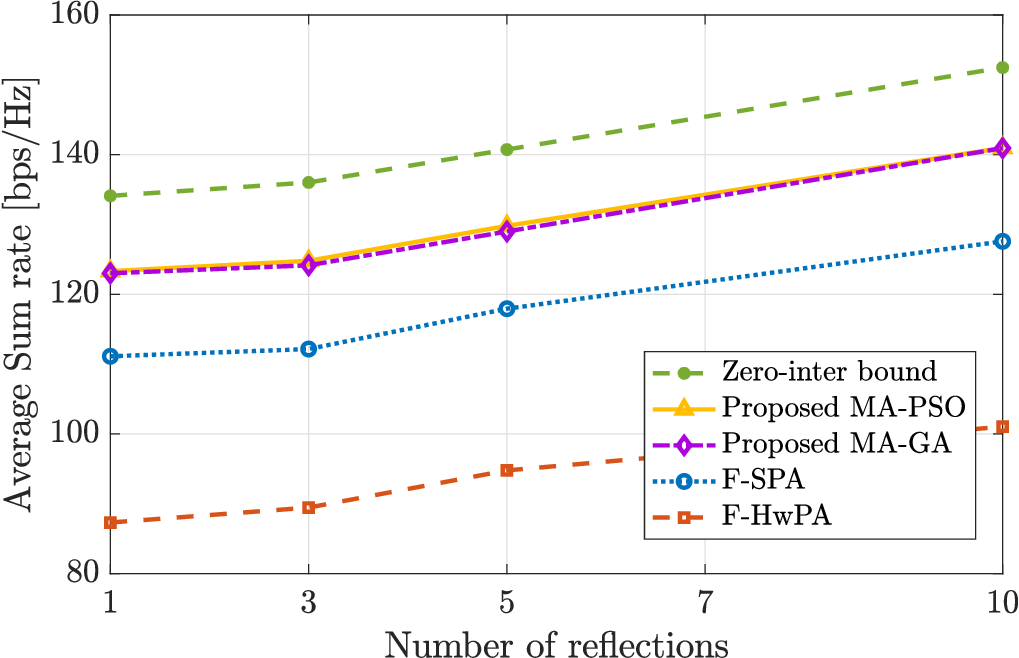}
    \caption{Average sum rate for varying numbers of reflections in the ray-tracing simulation.}
    \label{fig:result_numRD}
\end{figure}

 In the ray-tracing simulations, since a single diffraction is generally sufficient to capture typical urban propagation effects, the number of diffractions is fixed to one, while the number of reflections is varied to assess their impact on system performance. The corresponding results are shown in Fig.~\ref{fig:result_numRD}. As the number of reflections increases, additional propagation paths become available, enabling richer spatial diversity and improving the achievable sum rate, particularly in multiuser MIMO scenarios. Unlike stochastic channel models, where increased scattering is often implicitly assumed, ray-tracing simulations explicitly capture the emergence of dominant specular paths. While the sum rate increases with the number of reflections, the growth is sublinear and governed by the finite geometry of the environment. In practice, saturation is expected once all dominant propagation paths are captured; the considered reflection range focuses on typical urban propagation conditions rather than asymptotic behavior.

\begin{figure}[t!]
    \centering
    \begin{subfigure}[b]{0.2\textwidth}
        \centering
        \includegraphics[width=\textwidth]{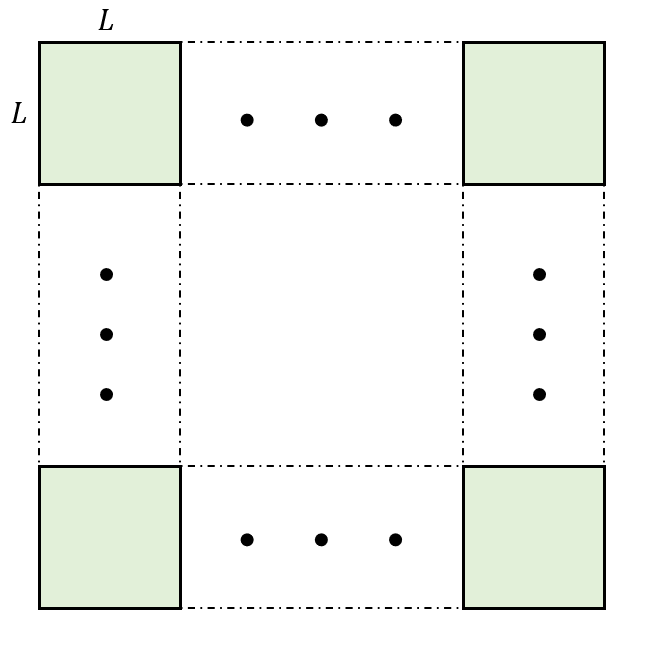}
        \caption{Varying array aperture size.}
        \label{fig:antAreaouter}
    \end{subfigure}
    \hspace{0.5cm}
    \begin{subfigure}[b]{0.2\textwidth}
        \centering
        \includegraphics[width=\textwidth]{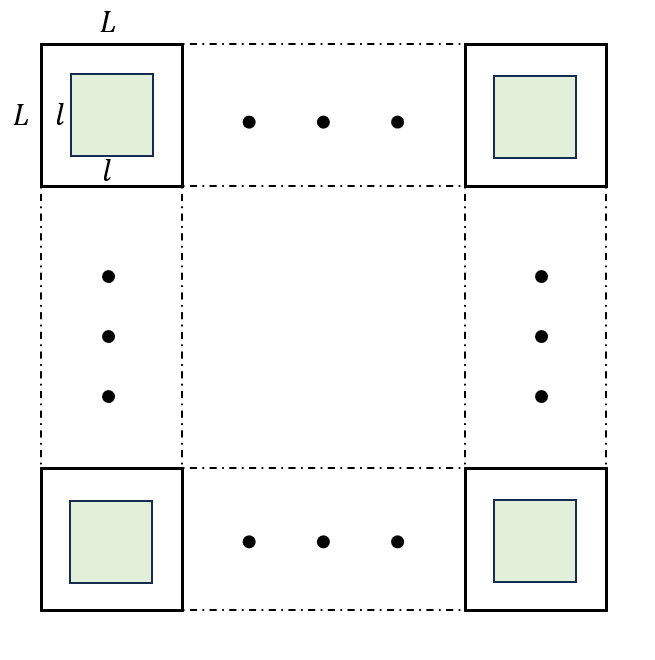}
        \caption{Varying antenna movement area (fixed aperture).}
        \label{fig:antAreainter}
    \end{subfigure}
    \caption{Two approaches for varying areas of movable antenna units.}
    \label{fig:result_AntennaRange}
\end{figure}
 
\begin{figure}[t!]
    \centering
    \begin{subfigure}[b]{0.241\textwidth}
        \centering
        \includegraphics[width=\textwidth]{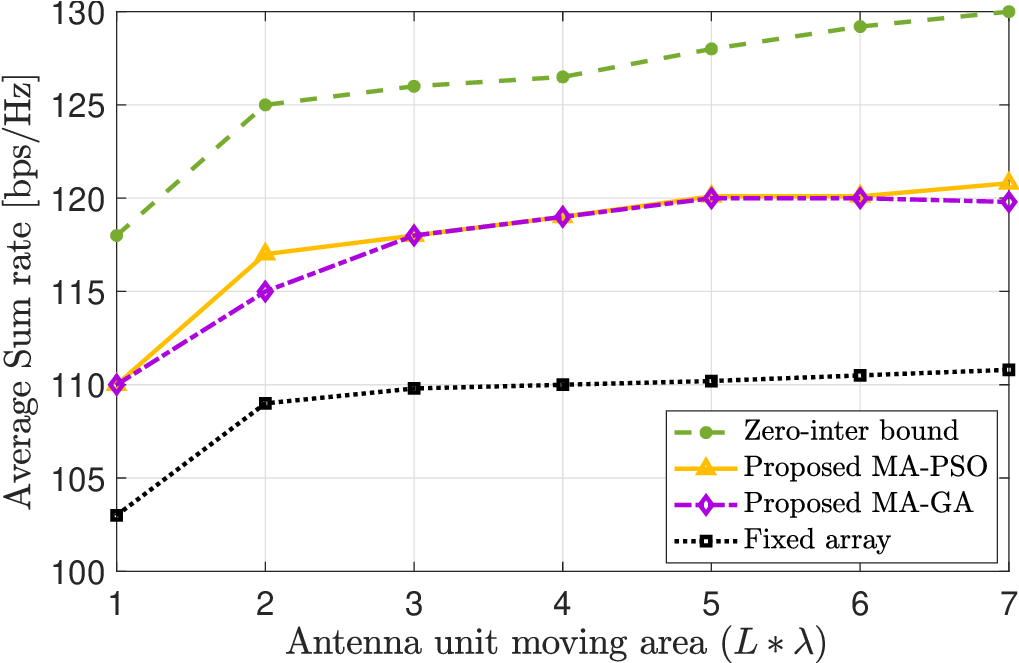}
        
        \caption{Average sum rate for varying antenna aperture.}
        \label{fig:sumrate_antAreaouter}
    \end{subfigure}
    \begin{subfigure}[b]{0.241\textwidth}
        \centering
        \includegraphics[width=\textwidth]{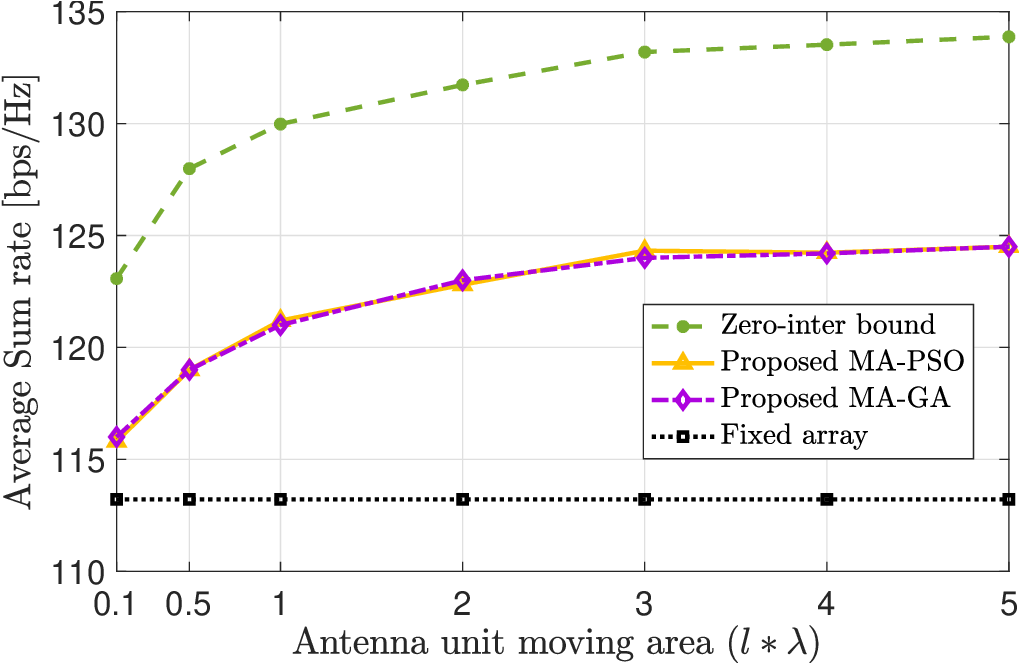}
        
        \caption{Average sum rate for varying MA constrained region.}
        \label{fig:sumrate_antAreainter}
    \end{subfigure}
    \caption{Sum rate comparison for different antenna unit areas.}
    \label{fig:sumrate_diff_area}
\end{figure}

 Fig.~\ref{fig:result_AntennaRange} illustrates the effect of changing array aperture and antenna movement range, where the area where each antenna can move within is represented by a green square. Fig.~\ref{fig:antAreaouter} shows the first case where increasing the side lengths of an antenna $L$ expands the array aperture. Fig.~\ref{fig:antAreainter} demonstrates the second scenario where the array aperture remains fixed ($5\lambda$) while the antenna movement area varies (where $l$ is varied from $0.1\lambda$ to $5\lambda$). It is observed that the average system sum rate increases monotonically with increasing antenna aperture, thereby enhancing spatial resolution and beamforming capability. A larger aperture enables the formation of narrower main lobes and improves the ability to resolve closely spaced multipath components and users, thereby enhancing spatial multiplexing performance. Although larger uniform inter-element spacings may introduce grating lobes when exceeding $\lambda/2$ the results indicate that the performance gains from increased aperture outweigh these effects in the considered environment as shown in Fig.~\ref{fig:sumrate_diff_area}(a).

In the second scenario, Fig.~\ref{fig:sumrate_diff_area}(b), as the movable range $l$ increases, the average system sum rate consistently improves. This behavior highlights a fundamental advantage of movable antenna systems: increased spatial flexibility allows the optimization algorithm to explore a larger configuration space and identify antenna placements that better exploit the instantaneous propagation environment. These results indicate that movable antenna systems are particularly attractive for deployments with constrained array aperture, where controlled antenna repositioning within the available space can achieve gains that would otherwise require a larger fixed array. 

\section{Conclusion}

This paper investigated downlink multiuser MIMO systems employing movable antennas under realistic propagation conditions modeled using deterministic ray-tracing techniques. The central question addressed was whether the performance gains reported for movable antenna systems under simplified distance-based channel models persist in geometry-constrained, three-dimensional propagation environments.
Simulation results demonstrate that movable antenna systems consistently outperform conventional fixed-array configurations even with realistic ray-traced channels with limited multipath richness, confirming that MA gains do not rely on artificially rich scattering assumptions. Moreover, a fundamental trade-off between antenna mobility and array aperture was identified, showing that while large fixed arrays can partially compensate for the lack of mobility, MA systems provide substantial benefits in aperture-constrained scenarios by exploiting dominant specular propagation paths.
From an algorithmic perspective, both particle swarm optimization and genetic algorithms achieve comparable performance, with GA exhibiting lower runtime and PSO offering greater parameter flexibility. These results provide new insights into the practical performance limits of movable antenna systems and clarify their role as a flexible array architecture for future wireless networks. Extending the proposed framework to dynamic user mobility and time-varying propagation environments constitutes an interesting direction for future work.

\bibliographystyle{IEEEtran}
\bibliography{references}

\end{document}